\documentclass[12pt]{article}

\begin{document}

\title{Perturbations and Stability of Black Ellipsoids }
\author{Sergiu I. Vacaru \thanks{%
E-mail address:\ vacaru@fisica.ist.utl.pt, ~~ sergiu$_{-}$vacaru@yahoo.com,\
} \\
{\small \textit{Centro Multidisciplinar de Astrofisica - CENTRA,
Departamento de Fisica,}}\\
{\small \textit{Instituto Superior Tecnico, Av. Rovisco Pais 1, Lisboa,
1049-001, Portugal}}\\
}
\date{July 21, 2003}
\maketitle

\begin{abstract}
We study the perturbations of two classes of static black ellipsoid
solutions of four dimensional vacuum Einstein equations. Such solutions are
described by generic off--diagonal metrics which are generated by
anholonomic transforms of diagonal metrics. The analysis is performed in the
approximation of small eccentricity deformations of the Schwarzschild
solution. We conclude that such anisotropic black hole objects may be stable
with respect to the perturbations parametrized by the Schrodinger equations
in the framework of the one--dimensional inverse scattering theory.

\vskip5pt.

Pacs 04.20.Jb,\ 04.70.-s,\ 04.70.Bw

\end{abstract}



\section{Introduction}

In a series of works new classes of solutions of four dimensional (4D) and
5D vacuum Einstein equations with ellipsoid and toroidal symmetry are
constructed \cite{v}. Such solutions are generated by anholonomic
deformations of the Schwarzschild metric,  of static or stationary
configurations, and depend anisotropically on angular coordinates. They are
described by generic off--diagonal metric ansatz which are effectively
diagonalized with respect to anholonomic frames with associated nonlinear
connection structure. A study of horizons and geodesic behaviour \cite{v1,v2}
concluded that for small deformations of the spherical symmetry, for
instance, to a resolution ellipsoid one. Such solutions define black
ellipsoid objects (static black holes with ellipsoidal horizons and
anisotropic polarizations of constants). With respect to anholonomic frames
the new solutions are given by certain metric coefficients which are similar
to those from the Reissner--Nordstrom metric, but with an effective,
polarized, ''electromagnetic'' charge, induced by off--diagonal vacuum
gravitational interactions. This  differs substantially from the usual
static electrovacuum solutions with spherical symmetry which are exact
solutions of the Einstein--Maxwell equations (see, for instance Ref.  \cite%
{chan}, as a general reference on ''the mathematical theory of black
holes'').

The aim of this paper is to study perturbations of black ellipsoids and to
prove that there are such static ellipsoid like configurations which are
stable with respect to perturbations of a fixed type of anisotropy (i. e.
for certain imposed anholonomic constraints). The main idea of a such proof
is to consider small (ellipsoidal, or another type) deformations of the
Schwarzschild metric and than to apply the already developed methods of the
theory of perturbations of classical black hole solutions, with a
re--definition of the formalism for adapted anholonomic frames.

The theory of perturbations of the Schwarzschild spacetime black holes was
initiated in Ref. \cite{rw}, developed in a series of works, e. g. Refs \cite%
{vis,fried}, and related \cite{dei} to the theory of inverse scattering and
its ramifications (see, for instanse, Refs. \cite{fad}). The results on the
theory of perturbations and stability of the Schwarzschild,
Reissner--Nordstrom and Kerr solutions are summarized in a monograph \cite%
{chan}. As alternative treatments of the stability of black holes we cite in
Ref. \cite{mon}.

The paper has the following structure: In Sec. 2 we introduce an
off--diagon\-al ansatz which parametrizes two classes of anholonomic static
deformations of the Schwar\-z\-schild solutions which describe black
ellipsoid like objects (the formulas for the components of Ricci and
Einstein tensors are outlined in the Appendix). In Sec. 3 we investigate the
axial metric perturbations governed by some one--dimensional Schrodinger
equations with nonlinear potential and anisotropic gravitational
polarizations. Section 4 is devoted to a stability analysis of polar metric
perturbations; a procedure of definition of formal solutions for polar
perturbations is formulated. In Sec. 5 we prove the stability of static
anholonomically deformed solutions of the Scwarschild metric with respect to
perturbations treated in the framework of the inverse scattering theory
based on the one--dimensional Schrodinger \ equation with paremetric
potentials. A discussion and conclusions are contained in Sec. 6.

\section{Metrics Describing Perturbations of Anisotropic Black Holes}

We consider a four dimensional pseudo--Riemannian quadratic linear element
\begin{eqnarray}
ds^{2} &=&\Omega (r,\varphi )\left[ -\left( 1-\frac{2m}{r}+\frac{\varepsilon
}{r^{2}}\right) ^{-1}dr^{2}-r^{2}q(r)d\theta ^{2}-\eta _{3}(r,\theta
,\varphi )r^{2}\sin ^{2}\theta \delta \varphi ^{2}\right]   \label{metric} \\
&&+\left[ 1-\frac{2m}{r}+\frac{\varepsilon }{r^{2}}\eta (r,\varphi )\right]
\delta t^{2},  \nonumber
\end{eqnarray}%
with
\[
\delta \varphi =d\varphi +\varepsilon w_{1}(r,\varphi )dr,\mbox{ and }\delta
t=dt+\varepsilon n_{1}(r,\varphi )dr,
\]%
where the local coordinates are denoted $u=\{u^{\alpha }=\left( r,\theta
,\varphi ,t\right) \}$ (the Greek indices $\alpha ,\beta ,...$ will run the
values 1,2,3,4), $\varepsilon $ is a small parameter satisfying the
conditions $0\leq \varepsilon \ll 1$ (for instance, an eccentricity for some
ellipsoid deformations of the spherical symmetry) and the functions $\Omega
(r,\varphi ),q(r),$ $\eta _{3}(r,\theta ,\varphi )$ and $\eta (\theta
,\varphi )$ are of necessary smooth class. The metric (\ref{metric}) is
static, off--diagonal and transforms into the usual Schwarzschild solution
if $\varepsilon \rightarrow 0$ and $q,\eta _{3}\rightarrow 1;$ it describes
at least two classes of static black hole solutions generated as small
anhlonomic deformations of the Schwarzschild solution \cite{v,v1,v2}.

The geodesic and the horizon structure of the first class of solutions was
investigated in Ref. \cite{v1}. It was proved that for the data
\begin{eqnarray*}
\eta _{3}(r,\varphi ) &=&\eta _{3[0]}(r,\varphi )+\varepsilon \lambda
_{3}(r,\varphi )+\varepsilon ^{2}\gamma _{3}(r,\varphi )+..., \\
\eta _{4}(r,\varphi ) &=&1+\varepsilon \lambda _{4}(r,\varphi )-\varepsilon
^{2}\lambda _{4}(r,\varphi )\left( 1-\frac{2m}{r}\right) ^{-1}+..., \\
\eta (r,\varphi ) &=&\lambda _{4}(r,\varphi )\left( r^{2}-2mr\right)
+1,\Omega =1, \\
\varepsilon \sqrt{|\eta _{3}|} &=&\eta _{0}\partial \sqrt{|\eta _{4}|}%
/\partial \varphi ,\eta _{0}=const,w_{1}=0,
\end{eqnarray*}%
and
\begin{eqnarray}
n_{1}(r,\varphi ) &=&n_{1[1]}\left( r\right) +n_{1[2]}\left( r\right) \int
d\varphi \ \eta _{3}\left( r,\varphi \right) /\left( \sqrt{|\eta _{4}\left(
\xi ,\varphi \right) |}\right) ^{3},\eta _{4}^{\ast }\neq 0;  \nonumber \\
&=&n_{1[1]}\left( r\right) +n_{1[2]}\left( r\right) \int d\varphi \ \eta
_{3}\left( r,\varphi \right) ,\eta _{4}^{\ast }=0;  \label{n1a} \\
&=&n_{1[1]}\left( r\right) +n_{1[2]}\left( r\right) \int d\varphi /\left(
\sqrt{|\eta _{4}\left( r,\varphi \right) |}\right) ^{3},\eta _{3}^{\ast }=0;
\nonumber
\end{eqnarray}%
where the limit $\partial \sqrt{|\eta _{4}|}/\partial \varphi \rightarrow 0$
is considered for $\varepsilon \rightarrow 0$ and the functions $\eta
_{3}\left( r,\varphi \right) $ and $n_{1[1,2]}\left( r\right) $ are stated
by some boundary conditions, there is defined a static black hole with
ellipsoidal horizon (a black ellipsoid).

Another class of black ellipsoids with anisotropic conformal symmetries,
with a nontrivial conformal factor $\Omega $ which induces a nonzero value
for the coefficient $w_{1}$ (see details in Ref. \cite{v2}), can be defined
by a metric (\ref{metric}) $\ $with the data
\begin{eqnarray*}
\Omega (r,\varphi ) &=&1-\varepsilon r^{2}\sin ^{2}\theta \eta _{3}(r,\theta
,\varphi ); \\
\eta (r,\varphi ) &=&r\eta _{3}(r,\theta ,\varphi )\sin ^{2}\theta \left(
r^{2}-2mr\right) ^{2}/2m; \\
\epsilon w_{1}(r,\varphi ) &=&\partial _{1}\Omega (\partial \Omega /\partial
\varphi )^{-1},\partial \eta _{3}/\partial \varphi \neq 0, \\
&=&0,\partial \eta _{3}/\partial \varphi =0;
\end{eqnarray*}%
and
\begin{equation}
n_{1}(r,\varphi )=n_{1[1]}\left( r\right) +n_{1[2]}\left( r\right) \int
d\varphi \ \eta _{3}\left( r,\varphi \right) ,  \label{n1b}
\end{equation}%
where, for this type of solutions, the function $\eta _{3}\left( r,\theta
,\varphi \right) $ is chosen as to have a value $\phi (r,\varphi )=\eta
_{3}\left( r,\theta ,\varphi \right) \sin ^{2}\theta $ depending only on
variables $r$ and $\varphi .$

The maximal analytic extensions of such locally anisotropic black hole
metrics in the framework of general relativity theory with anholonomic
frames were constructed in Refs. \cite{v1,v2}. It was shown that the
condition of vanishing of the coefficient $1-2m/r+\varepsilon \eta
(r,\varphi )/r^{2}$ before $\delta t$ defines the equation of a static
non--spherical horizon for a so called ''locally anisotropic'' black hole
(for a corresponding parametrization of $\eta $ we may construct a
resolution ellipsoid horizon). There is a similarity of the metrics of type (%
\ref{metric}) with the Reissner--Norstrom solution: the anisotropic metric
is obtained as a linear approximation on $\varepsilon $ from some exact
vacuum solutions of the Einstein equations, but the Reissner--Norstrom one
is a static exact solution of the Einstein--Maxwell equations.

We can apply the perturbation theory for the metric (\ref{metric}) (not
paying a special attention to some particular parametrization of
coefficients for one or another class of anisotropic black hole solutions)
and analyze its stability by using the results of Ref. \cite{chan} for a
fixed anisotropic direction, i. e. by imposing certain anholonomic frame
constraints for an angle $\varphi =\varphi _{0}$ but considering possible
perturbations depending on three variables $(u^{1}=x^{1}=r,u^{2}=x^{2}=%
\theta ,$ $u^{4}=t).$ If we prove that there is a stability on perturbations
for a value $\varphi _{0},$ we can analyze in a similar manner another
values of $\varphi .$ A more general perturbative theory with variable
anisotropy on coordinate $\varphi ,$ i. e. with dynamical anholonomic
constraints, connects the approach with a two dimensional inverse problem
which makes the analysis more sophisticate.

We note that in a study of perturbations of any spherically symmetric system
\ and, for instance, of small ellipsoid deformations, without any loss of
generality, we can restrict our considerations to axisymmetric modes of
perturbations. Non--axisymmetric modes of perturbations with an $%
e^{in\varphi }$ dependence on the azimutal angle $\varphi $ $\ (n$ being an
integer number) can be deduced from modes of axisymmetric perturbations with
$n=0$ by suitable rotations since there are not preferred axes in a
spherically symmetric background. The ellipsoid like deformations may be
included into the formalism as some low frequency and constrained
petrurbations.

For simplicity, in this paper, we restrict our study only to fixed values of
the coordinate $\varphi $ assuming that anholonomic deformations are
proportional to a small parameter $\varepsilon ;$ we shall investigate the
stability of solutions only by applying the one dimensional inverse methods.
The metric (\ref{metric}) to be investigated is with a deformed horizon and
transforms into the usual Schwarszschild solution at long radial distances
(see Refs. \cite{v1,v2} for details on such off--diagonal metrics).

Let us consider a quadratic metric element
\begin{eqnarray}
ds^{2} &=&-e^{2\mu _{1}}(du^{1})^{2}-e^{2\mu _{2}}(du^{2})^{2}-e^{2\mu
_{3}}(\delta u^{3})^{2}+e^{2\mu _{4}}(\delta u^{4})^{2},  \nonumber \\
\delta u^{3} &=&d\varphi -q_{1}dx^{1}-q_{2}dx^{2}-\omega dt,  \label{metric2}
\\
\delta u^{4} &=&dt+n_{1}dr  \nonumber
\end{eqnarray}%
where
\begin{eqnarray}
\mu _{\alpha }(x^{k},t) &=&\mu _{\alpha }^{(\varepsilon )}(x^{k},\varphi
_{0})+\delta \mu _{\alpha }^{(\varsigma )}(x^{k},t),  \label{coef1} \\
q_{i}(x^{k},t) &=&q_{i}^{(\varepsilon )}(r,\varphi _{0})+\delta
q_{i}^{(\varsigma )}(x^{k},t),  \nonumber \\
\omega (x^{k},t) &=&0+\delta \omega ^{(\varsigma )}(x^{k},t)  \nonumber
\end{eqnarray}%
with
\begin{eqnarray}
e^{2\mu _{1}^{(\varepsilon )}} &=&\Omega (r,\varphi _{0})(1-\frac{2m}{r}+%
\frac{\varepsilon }{r^{2}})^{-1},  \label{coef2} \\
e^{2\mu _{2}^{(\varepsilon )}} &=&\Omega (r,\varphi _{0})r^{2},  \nonumber \\
e^{2\mu _{3}^{(\varepsilon )}} &=&\Omega (r,\varphi _{0})r^{2}\sin
^{2}\theta \eta _{3}(r,\varphi _{0}),  \nonumber \\
e^{2\mu _{4}^{(\varepsilon )}} &=&1-\frac{2m}{r}+\frac{\varepsilon }{r^{2}}%
\eta (r,\varphi _{0}),  \nonumber
\end{eqnarray}%
and some non--trivial values for $q_{i}^{(\varepsilon )}$ and $\varepsilon
n_{i},$%
\begin{eqnarray*}
q_{1}^{(\varepsilon )} &=&\varepsilon w_{1}(r,\varphi _{0}), \\
n_{1} &=&\varepsilon \left( n_{1[1]}(r)+n_{1[2]}(r)\int_{0}^{\varphi
_{0}}\eta _{3}(r,\varphi )d\varphi \right) .
\end{eqnarray*}

We are distinguishing two types of small deformations from the spherical
symmetry. The first type of deformations, labeled with the index $%
(\varepsilon )$ are generated by some $\varepsilon $--terms which define a
fixed ellipsoid like configuration and the second type ones, labeled with
the index $(\varsigma ),$ are some small linear fluctuations of the metric
coefficients

The general formluas for the Ricci and Einstein tensors for metric elements
of class (\ref{metric2}) with $n_{1}=0$ are given in \cite{chan}. We compute
similar values with respect to anholnomic frames, when, for a conventional
splitting $u^{\alpha }=(x^{i},y^{a}),$ the coordinates $x^{i}$ and $y^{a}$
are treated respectively as holonomic and anholonomic ones. In this case the
partial derivatives $\partial /\partial x^{i}$ must be changed into certain
'elongated' ones
\begin{eqnarray*}
\frac{\partial }{\partial x^{1}} &\rightarrow &\frac{\delta }{\partial x^{1}}%
=\frac{\partial }{\partial x^{1}}-w_{1}\frac{\partial }{\partial \varphi }%
-n_{1}\frac{\partial }{\partial t}, \\
\frac{\partial }{\partial x^{2}} &\rightarrow &\frac{\delta }{\partial x^{1}}%
=\frac{\partial }{\partial x^{1}}-w_{2}\frac{\partial }{\partial \varphi }%
-n_{2}\frac{\partial }{\partial t},
\end{eqnarray*}%
see details in Refs \cite{v,v1,v2}. In the ansatz (\ref{metric2}), the
anholonomic contributions of $w_{i}$ are included in the coefficients $%
q_{i}(x^{k},t),$ there is only one \ nonzero value $w_{1}$ and, in
consequence, we have to introduce elongations of partial space derivatives
only on the $t$--variable. For convenience, in the Appendix we present the
necessary formulas for $R_{\alpha \beta }$ (the Ricci tensor) and $G_{\alpha
\beta }$ (the Einstein tensor) computed for the ansatz (\ref{metric2}) with
three holonomic coordinates $\left( r,\theta ,\varphi \right) $ and on
anholonomic coodinate $t$ (in our case, being time like), with the partial
derivative operators
\[
\partial _{1}\rightarrow \delta _{1}=\frac{\partial }{\partial r}-n_{1}\frac{%
\partial }{\partial t},\partial _{2}=\frac{\partial }{\partial \theta }%
,\partial _{3}=\frac{\partial }{\partial \varphi },
\]%
and for a fixed value $\varphi _{0}.$

A general perturbation of an anisotropic black--hole described by a
quadratic line element (\ref{metric2}) \ results in some small quantities of
the first order $\omega $ and $q_{i},$ inducing a dragging of frames and
imparting rotations, and in some functions $\mu _{\alpha }$ with small
increments $\delta \mu _{\alpha },$ which do not impart rotations. Some
coefficients contained in such values are proportional to $\varepsilon ,$
another ones are considered only as small quantities. The perturbations of
metric are of two generic types: axial and polar one. We investigate them
separately in the next two Sections.

\section{Axial metric perturbations}

Axial perturbations are characterized by non--vanishing $\omega $ and $q_{i}$
which satisfy the equations
\[
R_{3i}=0,
\]%
see the explicit formulas for such coefficients of the Ricci tensor in the
Appendix. The resulting equations governing axial perturbations, $\delta
R_{31}=0,$ $\delta R_{32}=0,$ are respectively%
\begin{eqnarray}
\partial _{2}\left( e^{3\mu _{3}^{(\varepsilon )}+\mu _{4}^{(\varepsilon
)}-\mu _{1}^{(\varepsilon )}-\mu _{2}^{(\varepsilon )}}Q_{12}\right)
&=&-e^{3\mu _{3}^{(\varepsilon )}-\mu _{4}^{(\varepsilon )}-\mu
_{1}^{(\varepsilon )}+\mu _{2}^{(\varepsilon )}}\partial _{4}Q_{14},
\label{eq1} \\
\delta _{1}\left( e^{3\mu _{3}^{(\varepsilon )}+\mu _{4}^{(\varepsilon
)}-\mu _{1}^{(\varepsilon )}-\mu _{2}^{(\varepsilon )}}Q_{12}\right)
&=&e^{3\mu _{3}^{(\varepsilon )}-\mu _{4}^{(\varepsilon )}+\mu
_{1}^{(\varepsilon )}-\mu _{2}^{(\varepsilon )}}\partial _{4}Q_{24},
\nonumber
\end{eqnarray}%
where
\begin{equation}
Q_{ij}=\delta _{i}q_{j}-\delta _{j}q_{i},Q_{i4}=\partial _{4}q_{i}-\delta
_{i}\omega  \label{eq1a}
\end{equation}%
and for $\mu _{i}$ there are considered unperturbed values $\mu
_{i}^{(\varepsilon )}.$ Introducing the values of coefficients (\ref{coef1})
and (\ref{coef2}) \ and assuming that the perturbations have a time
dependence of type $\exp (i\sigma t)$ for a real constant $\sigma ,$ \ we
rewrite the equations (\ref{eq1})
\begin{eqnarray}
\frac{1+\varepsilon \left( \Delta ^{-1}+3r^{2}\phi /2\right) }{r^{4}\sin
^{3}\theta \eta _{3}^{3/2}}\partial _{2}Q^{(\eta )} &=&-i\sigma \delta
_{r}\omega -\sigma ^{2}q_{1},  \label{eq2a} \\
\frac{\Delta }{r^{4}\sin ^{3}\theta \eta _{3}^{3/2}}\delta _{1}\left\{
Q^{(\eta )}\left[ 1+\frac{\varepsilon }{2}\left( \frac{\eta -1}{\Delta }%
-r^{2}\phi \right) \right] \right\} &=&i\sigma \partial _{\theta }\omega
+\sigma ^{2}q_{2}  \label{eq2b}
\end{eqnarray}%
for
\[
Q^{(\eta )}(r,\theta ,\varphi _{0},t)=\Delta Q_{12}\sin ^{3}\theta =\Delta
\sin ^{3}\theta (\partial _{2}q_{1}-\delta _{1}q_{2}),\Delta =r^{2}-2mr,
\]%
where $\phi =0$ for solutions with $\Omega =1$ and $\phi (r,\varphi )=\eta
_{3}\left( r,\theta ,\varphi \right) \sin ^{2}\theta ,$ i. e. $\eta
_{3}\left( r,\theta ,\varphi \right) \sim \sin ^{-2}\theta $ for solutions
with $\Omega =1+\varepsilon ....$

We can exclude the function $\omega $ and define an equation for $Q^{(\eta
)} $ if we take the sum of the (\ref{eq2a}) subjected by the action of
operator $\partial _{2}$ and of the (\ref{eq2b}) subjected by the action of
operator $\delta _{1}.$ Using the relations (\ref{eq1a}), we write%
\begin{eqnarray*}
r^{4}\delta _{1}\left\{ \frac{\Delta }{r^{4}\eta _{3}^{3/2}}\left[ \delta
_{1}\left[ Q^{(\eta )}+\frac{\varepsilon }{2}\left( \frac{\eta -1}{\Delta }%
-r^{2}\right) \phi \right] \right] \right\} + && \\
\sin ^{3}\theta \partial _{2}\left[ \frac{1+\varepsilon (\Delta
^{-1}+3r^{2}\phi /2)}{\sin ^{3}\theta \eta _{3}^{3/2}}\partial _{2}Q^{(\eta
)}\right] +\frac{\sigma ^{2}r^{4}}{\Delta \eta _{3}^{3/2}}Q^{(\eta )} &=&0.
\end{eqnarray*}%
The solution of this equation is searched in the form $Q^{(\eta
)}=Q+\varepsilon Q^{(1)}$ which results in
\begin{equation}
r^{4}\partial _{1}\left( \frac{\Delta }{r^{4}\eta _{3}^{3/2}}\partial
_{1}Q\right) +\sin ^{3}\theta \partial _{2}\left( \frac{1}{\sin ^{3}\theta
\eta _{3}^{3/2}}\partial _{2}Q\right) +\frac{\sigma ^{2}r^{4}}{\Delta \eta
_{3}^{3/2}}Q=\varepsilon A\left( r,\theta ,\varphi _{0}\right) ,  \label{eq3}
\end{equation}%
where%
\begin{eqnarray*}
A\left( r,\theta ,\varphi _{0}\right) &=&r^{4}\partial _{1}\left( \frac{%
\Delta }{r^{4}\eta _{3}^{3/2}}n_{1}\right) \frac{\partial Q}{\partial t}%
-r^{4}\partial _{1}\left( \frac{\Delta }{r^{4}\eta _{3}^{3/2}}\partial
_{1}Q^{(1)}\right) \\
&&-\sin ^{3}\theta \partial _{2}\left[ \frac{1+\varepsilon (\Delta
^{-1}+3r^{2}\phi /2)}{\sin ^{3}\theta \eta _{3}^{3/2}}\partial _{2}Q^{(1)}-%
\frac{\sigma ^{2}r^{4}}{\Delta \eta _{3}^{3/2}}Q^{(1)}\right] ,
\end{eqnarray*}%
with a time dependence like $\exp [i\sigma t]$

It is possible to construct different classes of solutions of the equation (%
\ref{eq3}). At the first step we find the solution for $Q$ when $\varepsilon
=0.$ Then, for a known value of $Q\left( r,\theta ,\varphi _{0}\right) $
from
\[
Q^{(\eta )}=Q+\varepsilon Q^{(1)},
\]%
we can define $Q^{(1)}$ from the equations (\ref{eq2a}) and (\ref{eq2b}) by
considering the values proportional to $\varepsilon $ which can be written
\begin{eqnarray}
\partial _{1}Q^{(1)} &=&B_{1}\left( r,\theta ,\varphi _{0}\right) ,
\label{eq4} \\
\partial _{2}Q^{(1)} &=&B_{2}\left( r,\theta ,\varphi _{0}\right) ,
\nonumber
\end{eqnarray}%
where%
\[
B_{1}=n_{1}\frac{\partial Q}{\partial t}-\frac{1}{2}\partial _{1}\left[
Q\left( \frac{\eta -1}{\Delta }-r^{2}\phi \right) \right]
\]%
and
\[
B_{2}=-(\Delta ^{-1}+3r^{2}\phi /2)\partial _{2}Q-\sigma ^{2}r^{4}\sin
^{3}\theta ~w_{1}.
\]%
The integrability condition of the system (\ref{eq4}), $\partial
_{1}B_{2}=\partial _{2}B_{1}$ imposes a relation between the polarization
functions $\eta _{3},\eta ,w_{1}$and $n_{1}$ (for a corresponding class of
solutions, see formulas (\ref{n1a}), or (\ref{n1b})). In order to prove that
there are stable anisotropic configurations of anisotropic black hole
solutions, we may consider a set of polarization functions when $A\left(
r,\theta ,\varphi _{0}\right) =0$ and the solution with $Q^{(1)}=0$ is
admitted. This holds, for example, if
\[
\Delta n_{1}=n_{0}r^{4}\eta _{3}^{3/2},\ n_{0}=const.
\]%
In this case the axial perturbations are described by the equation
\begin{equation}
\eta _{3}^{3/2}r^{4}\partial _{1}\left( \frac{\Delta }{r^{4}\eta _{3}^{3/2}}%
\partial _{1}Q\right) +\sin ^{3}\theta \partial _{2}\left( \frac{1}{\sin
^{3}\theta }\partial _{2}Q\right) +\frac{\sigma ^{2}r^{4}}{\Delta }Q=0
\label{eq5}
\end{equation}%
which is obtained from (\ref{eq3}) for $\eta _{3}=\eta _{3}\left( r,\varphi
_{0}\right) ,$ or for $\phi (r,\varphi _{0})=\eta _{3}\left( r,\theta
,\varphi _{0}\right) \sin ^{2}\theta .$

In the limit $\eta _{3}\rightarrow 1$ the solution of equation (\ref{eq5})
is investigated in details in Ref. \cite{chan}. \ Here, we prove that in a
similar manner we can define exact solutions for non--trivial values of $%
\eta _{3}.$ \ The variables $r$ and $\theta $ can be separated if we
substitute
\[
Q(r,\theta ,\varphi _{0})=Q_{0}(r,\varphi _{0})C_{l+2}^{-3/2}(\theta ),
\]%
where $C_{n}^{\nu }$ are the Gegenbauer functions generated by the equation%
\[
\left[ \frac{d}{d\theta }\sin ^{2\nu }\theta \frac{d}{d\theta }+n\left(
n+2\nu \right) \sin ^{2\nu }\theta \right] C_{n}^{\nu }(\theta )=0.
\]%
The function $C_{l+2}^{-3/2}(\theta )$ is related to the second derivative
of the Legendre function $P_{l}(\theta )$ by formulas%
\[
C_{l+2}^{-3/2}(\theta )=\sin ^{3}\theta \frac{d}{d\theta }\left[ \frac{1}{%
\sin \theta }\frac{dP_{l}(\theta )}{d\theta }\right] .
\]%
The separated part of (\ref{eq5}) depending on radial variable with a fixed
value $\varphi _{0}$ transforms into the equation%
\begin{equation}
\eta _{3}^{3/2}\Delta \frac{d}{dr}\left( \frac{\Delta }{r^{4}\eta _{3}^{3/2}}%
\frac{dQ_{0}}{dr}\right) +\left( \sigma ^{2}-\frac{\mu ^{2}\Delta }{r^{4}}%
\right) Q_{0}=0,  \label{eq6}
\end{equation}%
where $\mu ^{2}=(l-1)(l+2)$ for $l=2,3,...$ A further simplification is
possible for $\eta _{3}=\eta _{3}(r,\varphi _{0})$ if we introduce in the
equation (\ref{eq6}) a new radial coordinate
\[
r_{\#}=\int \eta _{3}^{3/2}(r,\varphi _{0})r^{2}dr
\]%
and a new unknown function $Z^{(\eta )}=r^{-1}Q_{0}(r).$ The equation for $%
Z^{(\eta )}$ is an Schrodinger like one--dimensional wave equation%
\begin{equation}
\left( \frac{d^{2}}{dr_{\#}^{2}}+\frac{\sigma ^{2}}{\eta _{3}^{3/2}}\right)
Z^{(\eta )}=V^{(\eta )}Z^{(\eta )}  \label{eq7}
\end{equation}%
with the potential
\begin{equation}
V^{(\eta )}=\frac{\Delta }{r^{5}\eta _{3}^{3/2}}\left[ \mu ^{2}-r^{4}\frac{d%
}{dr}\left( \frac{\Delta }{r^{4}\eta _{3}^{3/2}}\right) \right]  \label{eq7a}
\end{equation}%
and polarized parameter
\[
\widetilde{\sigma }^{2}=\sigma ^{2}/\eta _{3}^{3/2}.
\]%
This equation transforms into the so--called Regge--Wheeler equation if $%
\eta _{3}=1.$ For instance, for the Schwarzschild black hole such solutions
are investigated and tabulated for different values of $l=2,3$ and $4$ in
Ref. \cite{chan}.

We note that for static anisotropic black holes with nontrivial anisotropic
conformal factor, $\Omega =1+\varepsilon ...,$ even $\eta _{3}$ may depend
on angular variable $\theta $ because of condition that $\phi (r,\varphi
_{0})=\eta _{3}\left( r,\theta ,\varphi _{0}\right) \sin ^{2}\theta $ the
equation (\ref{eq5}) transforms directly in (\ref{eq7}) with $\mu =0$
without any separation of variables $r$ and $\theta .$ It is not necessary
in this case to consider the Gegenbauer functions because $Q_{0}$ does not
depend on $\theta $ which corresponds to a solution with $l=1.$

We may transform (\ref{eq7}) into the usual form,
\[
\left( \frac{d^{2}}{dr_{\star }^{2}}+\sigma ^{2}\right) Z^{(\eta )}=%
\widetilde{V}^{(\eta )}Z^{(\eta )}
\]%
if we introduce the variable
\[
r_{\star }=\int dr_{\#}\eta _{3}^{-3/2}\left( r_{\#},\varphi _{0}\right)
\]%
for $\widetilde{V}^{(\eta )}=\eta _{3}^{3/2}V^{(\eta )}.$ So, the
polarization function $\eta _{3},$ describing static anholonomic
deformations of the Scharzshild black hole, ''renormalizes'' the potential
in the one--dimensional Schrodinger wave--equation governing axial
perturbations of such objects.

We conclude that small static ''ellipsoid'' like deformations and
polarizations of constants of spherical black holes (the anisotropic
configurations being described by generic vacuum off--diagonal metric
ansatz) do not change the type of equations for axial petrubations: one
modifies the potential barrier,%
\[
V^{(-)}=\frac{\Delta }{r^{5}}\left[ \left( \mu ^{2}+2\right) r-6m\right]
\longrightarrow \widetilde{V}^{(\eta )}
\]%
and re--defines the radial variables%
\[
r_{\ast }=r+2m\ln \left( r/2m-1\right) \longrightarrow r_{\star }(\varphi
_{0})
\]%
with a parametric dependence on anisotropic angular coordinate which is
caused by the existence of a deformed static horizon.

\section{Polar metric perturbations}

The polar perturbations are described by non--trivial increments of the
diagonal metric coefficients, $\delta \mu _{\alpha }=\delta \mu _{\alpha
}^{(\varepsilon )}+\delta \mu _{\alpha }^{(\varsigma )},$ for
\[
\mu _{\alpha }^{(\varepsilon )}=\nu _{\alpha }+\delta \mu _{\alpha
}^{(\varepsilon )}
\]%
where $\delta \mu _{\alpha }^{(\varsigma )}(x^{k},t)$ parametrize time
depending fluctuations which are stated to be the same both for spherical
and/or spheroid configurations and $\delta \mu _{\alpha }^{(\varepsilon )}$
is a static deformation from the spherical symmetry. Following notations (%
\ref{coef1}) and (\ref{coef2}) we write%
\[
e^{v_{1}}=r/\sqrt{|\Delta |},e^{v_{2}}=r\sqrt{|q(r)|},e^{v_{3}}=rh_{3}\sin
\theta ,e^{v_{4}}=\Delta /r^{2}
\]%
and
\[
\delta \mu _{1}^{(\varepsilon )}=-\frac{\varepsilon }{2}\left( \Delta
^{-1}+r^{2}\phi \right) ,\delta \mu _{2}^{(\varepsilon )}=\delta \mu
_{3}^{(\varepsilon )}=-\frac{\varepsilon }{2}r^{2}\phi ,\delta \mu
_{4}^{(\varepsilon )}=\frac{\varepsilon \eta }{2\Delta }
\]%
where $\phi =0$ for the solutions with $\Omega =1.$

Examining the expressions for $R_{4i},R_{12,}R_{33}$ and $G_{11}$ (see the
Appendix) we conclude that the values $Q_{ij}$ appear quadratically which
can be ignored in a linear perturbation theory. Thus the equations for the
axial and the polar perturbations decouple. Considering only linearized
expressions, both for static $\varepsilon $--terms and fluctuations
depending on time about the Schwarzschild values we obtain the equations%
\begin{eqnarray}
\delta _{1}\left( \delta \mu _{2}+\delta \mu _{3}\right) +\left(
r^{-1}-\delta _{1}\mu _{4}\right) \left( \delta \mu _{2}+\delta \mu
_{3}\right) -2r^{-1}\delta \mu _{1} &=&0\quad \left( \delta R_{41}=0\right) ,
\nonumber \\
\partial _{2}\left( \delta \mu _{1}+\delta \mu _{3}\right) +\left( \delta
\mu _{2}-\delta \mu _{3}\right) \cot \theta &=&0\quad \left( \delta
R_{42}=0\right) ,  \nonumber \\
\partial _{2}\delta _{1}\left( \delta \mu _{3}+\delta \mu _{4}\right)
-\delta _{1}\left( \delta \mu _{2}-\delta \mu _{3}\right) \cot \theta - &&
\nonumber \\
\left( r^{-1}-\delta _{1}\mu _{4}\right) \partial _{2}(\delta \mu
_{4})-\left( r^{-1}+\delta _{1}\mu _{4}\right) \partial _{2}(\delta \mu
_{1}) &=&0\quad \left( \delta R_{42}=0\right) ,  \nonumber \\
e^{2\mu _{4}}\{2\left( r^{-1}+\delta _{1}\mu _{4}\right) \delta _{1}(\delta
\mu _{3})+r^{-1}\delta _{1}\left( \delta \mu _{3}+\delta \mu _{4}-\delta \mu
_{1}+\delta \mu _{2}\right) + &&  \label{peq1} \\
\delta _{1}\left[ \delta _{1}(\delta \mu _{3})\right] -2r^{-1}\delta \mu
_{1}\left( r^{-1}+2\delta _{1}\mu _{4}\right) \}-2e^{-2\mu _{4}}\partial
_{4}[\partial _{4}(\delta \mu _{3})]+ &&  \nonumber \\
r^{-2}\{\partial _{2}[\partial _{2}(\delta \mu _{3})]+\partial _{2}\left(
2\delta \mu _{3}+\delta \mu _{4}+\delta \mu _{1}-\delta \mu _{2}\right) \cot
\theta +2\delta \mu _{2}\} &=&0\quad \left( \delta R_{33}=0\right) ,
\nonumber \\
e^{-2\mu _{1}}[r^{-1}\delta _{1}(\delta \mu _{4})+\left( r^{-1}+\delta
_{1}\mu _{4}\right) \delta _{1}\left( \delta \mu _{2}+\delta \mu _{3}\right)
- &&  \nonumber \\
2r^{-1}\delta \mu _{1}\left( r^{-1}+2\delta _{1}\mu _{4}\right) ]-e^{-2\mu
_{4}}\partial _{4}[\partial _{4}(\delta \mu _{3}+\delta \mu _{2})] &&
\nonumber \\
+r^{-2}\{\partial _{2}[\partial _{2}(\delta \mu _{3})]+\partial _{2}\left(
2\delta \mu _{3}+\delta \mu _{4}-\delta \mu _{2}\right) \cot \theta +2\delta
\mu _{2}\} &=&0\quad \left( \delta G_{11}=0\right) .  \nonumber
\end{eqnarray}

The values of type $\delta \mu _{\alpha }=\delta \mu _{\alpha
}^{(\varepsilon )}+\delta \mu _{\alpha }^{(\varsigma )}$ from (\ref{peq1})
contain two components: the first ones are static, proportional to $%
\varepsilon ,$ and the second ones may depend on time coordinate $t.$ We
shall assume that the perturbations $\delta \mu _{\alpha }^{(\varsigma )}$
have a time--dependence $\exp [\sigma t]$ $\ $so that the partial time
derivative $"\partial _{4}"$ is replaced by the factor $i\sigma .$ In order
to treat both type of increments in a similar fashion we may consider that
the values labeled with $(\varepsilon )$ also oscillate in time like $\exp
[\sigma ^{(\varepsilon )}t]$ but with a very small (almost zero) frequency $%
\sigma ^{(\varepsilon )}\rightarrow 0.$ There are also actions of
''elongated'' partial derivative operators like
\[
\delta _{1}\left( \delta \mu _{\alpha }\right) =\partial _{1}\left( \delta
\mu _{\alpha }\right) -\varepsilon n_{1}\partial _{4}\left( \delta \mu
_{\alpha }\right) .
\]%
To avoid a calculus with complex values we associate the terms proportional $%
\varepsilon n_{1}\partial _{4}$ to amplitudes of type $\varepsilon
in_{1}\partial _{4}$ and write this operator as
\[
\delta _{1}\left( \delta \mu _{\alpha }\right) =\partial _{1}\left( \delta
\mu _{\alpha }\right) +\varepsilon n_{1}\sigma \left( \delta \mu _{\alpha
}\right) .
\]%
For the ''non-perturbed'' Schwarzschild values, which are static, the
operator $\delta _{1}$ reduces to $\partial _{1},$ i.e. $\delta
_{1}v_{\alpha }=\partial _{1}v_{\alpha }.$ Hereafter we shall consider that
the solution of the system (\ref{peq1}) consists \ from a superposition of
two linear solutions, $\delta \mu _{\alpha }=\delta \mu _{\alpha
}^{(\varepsilon )}+\delta \mu _{\alpha }^{(\varsigma )}$; the first class of
solutions for increments will be provided with index $(\varepsilon ),$
corresponding to the frequence $\sigma ^{(\varepsilon )}$ and the second
class will be for the increments with index $(\varsigma )$ and correspond to
the frequence $\sigma ^{(\varsigma )}.$ We shall write this as $\delta \mu
_{\alpha }^{(A)}$ and $\sigma _{(A)}$ for the labels $A=\varepsilon $ or $%
\varsigma $ and suppress the factors $\exp [\sigma ^{(A)}t]$ in our
subsequent considerations. The system of equations (\ref{peq1}) will be
considered for both type of increments.

We can separate the variables by substitutions (see the method in Refs. \cite%
{fried,chan})%
\begin{eqnarray}
\delta \mu _{1}^{(A)} &=&L^{(A)}(r)P_{l}(\cos \theta ),\quad \delta \mu
_{2}^{(A)}=\left[ T^{(A)}(r)P_{l}(\cos \theta )+V^{(A)}(r)\partial
^{2}P_{l}/\partial \theta ^{2}\right] ,  \label{auxc1} \\
\delta \mu _{3}^{(A)} &=&\left[ T^{(A)}(r)P_{l}(\cos \theta )+V^{(A)}(r)\cot
\theta \partial P_{l}/\partial \theta \right] ,\quad \delta \mu
_{4}^{(A)}=N^{(A)}(r)P_{l}(\cos \theta )  \nonumber
\end{eqnarray}%
and reduce the system of equations (\ref{peq1}) \ to%
\begin{eqnarray}
\delta _{1}\left( N^{(A)}-L^{(A)}\right) &=&\left( r^{-1}-\partial _{1}\nu
_{4}\right) N^{(A)}+\left( r^{-1}+\partial _{1}\nu _{4}\right) L^{(A)},
\nonumber \\
\delta _{1}L^{(A)}+\left( 2r^{-1}-\partial _{1}\nu _{4}\right) N^{(A)} &=&-
\left[ \delta _{1}X^{(A)}+\left( r^{-1}-\partial _{1}\nu _{4}\right) X^{(A)}%
\right] ,  \label{peq2a}
\end{eqnarray}%
and
\begin{eqnarray}
2r^{-1}\delta _{1}\left( N^{(A)}\right)
-l(l+1)r^{-2}e^{-2v_{4}}N^{(A)}-2r^{-1}(r^{-1}+2\partial _{1}\nu
_{4})L^{(A)}-2(r^{-1}+ &&  \label{peq2b} \\
\partial _{1}\nu _{4})\delta _{1}\left[ N^{(A)}+(l-1)(l+2)V^{(A)}/2\right]
-(l-1)(l+2)r^{-2}e^{-2v_{4}}\left( V^{(A)}-L^{(A)}\right) - &&  \nonumber \\
2\sigma _{(A)}^{2}e^{-4v_{4}}\left[ L^{(A)}+(l-1)(l+2)V^{(A)}/2\right] &=&0,
\nonumber
\end{eqnarray}%
where we have introduced new functions
\[
X^{(A)}=\frac{1}{2}(l-1)(l+2)V^{(A)}
\]%
and considered the relation
\[
T^{(A)}-V^{(A)}+L^{(A)}=0\quad (\delta R_{42}=0).
\]%
We can introduce the functions
\begin{eqnarray}
\widetilde{L}^{(A)} &=&L^{(A)}+\varepsilon \sigma _{(A)}\int
n_{1}L^{(A)}dr,\quad \widetilde{N}^{(A)}=N^{(A)}+\varepsilon \sigma
_{(A)}\int n_{1}N^{(A)}dr,  \label{tilds} \\
\widetilde{T}^{(A)} &=&N^{(A)}+\varepsilon \sigma _{(A)}\int
n_{1}N^{(A)}dr,\quad \widetilde{V}^{(A)}=V^{(A)}+\varepsilon \sigma
_{(A)}\int n_{1}V^{(A)}dr,  \nonumber
\end{eqnarray}%
for which
\[
\partial _{1}\widetilde{L}^{(A)}=\delta _{1}\left( L^{(A)}\right) ,\partial
_{1}\widetilde{N}^{(A)}=\delta _{1}\left( N^{(A)}\right) ,\partial _{1}%
\widetilde{T}^{(A)}=\delta _{1}\left( T^{(A)}\right) ,\partial _{1}%
\widetilde{V}^{(A)}=\delta _{1}\left( V^{(A)}\right) ,
\]%
and, this way it is possible to substitute in (\ref{peq2a}) and (\ref{peq2b}%
) the elongated partial derivative $\delta _{1}$ by the usual one acting on
''tilded'' radial increments.

By straightforward calculations (see details in Ref. \cite{chan}) one can
check that the functions
\[
Z_{(A)}^{(+)}=r^{2}\frac{6mX^{(A)}/r(l-1)(l+2)-L^{(A)}}{r(l-1)(l+2)/2+3m}
\]%
satisfy one--dimensional wave equations similar to (\ref{eq7}) for $Z^{(\eta
)}$ with $\eta _{3}=1,$ when $r_{\star }=r_{\ast },$%
\begin{eqnarray}
\left( \frac{d^{2}}{dr_{\ast }^{2}}+\sigma _{(A)}^{2}\right) \widetilde{Z}%
_{(A)}^{(+)} &=&V^{(+)}Z_{(A)}^{(+)},  \label{peq3} \\
\widetilde{Z}_{(A)}^{(+)} &=&Z_{(A)}^{(+)}+\varepsilon \sigma _{(A)}\int
n_{1}Z_{(A)}^{(+)}dr,  \nonumber
\end{eqnarray}%
where%
\begin{eqnarray}
V^{(+)} &=&\frac{2\Delta }{r^{5}[r(l-1)(l+2)/2+3m]^{2}}\times \{9m^{2}\left[
\frac{r}{2}(l-1)(l+2)+m\right]  \label{pot3} \\
&&+\frac{1}{4}(l-1)^{2}(l+2)^{2}r^{3}\left[ 1+\frac{1}{2}(l-1)(l+2)+\frac{3m%
}{r}\right] \}.  \nonumber
\end{eqnarray}%
For $\varepsilon \rightarrow 0,$ the equation (\ref{peq3}) transforms in the
usual Zerilli equation \cite{zerilli,chan}.

To complete the solution we give the formulas for the ''tilded'' $L$--, $X$%
-- and $N$--factors,%
\begin{eqnarray}
\widetilde{L}^{(A)} &=&\frac{3m}{r^{2}}\widetilde{\Phi }^{(A)}-\frac{%
(l-1)(l+2)}{2r}\widetilde{Z}_{(A)}^{(+)},  \label{peg3a} \\
\widetilde{X}^{(A)} &=&\frac{(l-1)(l+2)}{2r}(\widetilde{\Phi }^{(A)}+%
\widetilde{Z}_{(A)}^{(+)}),  \nonumber \\
\widetilde{N}^{(A)} &=&\left( m-\frac{m^{2}+r^{4}\sigma _{(A)}^{2}}{r-2m}%
\right) \frac{\widetilde{\Phi }^{(A)}}{r^{2}}-\frac{(l-1)(l+2)r}{%
2(l-1)(l+2)+12m}\frac{\partial \widetilde{Z}_{(A)}^{(+)}}{\partial r_{\#}}
\nonumber \\
&&-\frac{(l-1)(l+2)}{\left[ r(l-1)(l+2)+6m\right] ^{2}}\times  \nonumber \\
&&\left\{ \frac{12m^{2}}{r}+3m(l-1)(l+2)+\frac{r}{2}(l-1)(l+2)\left[
(l-1)(l+2)+2\right] \right\} ,  \nonumber
\end{eqnarray}%
where
\[
\widetilde{\Phi }^{(A)}=(l-1)(l+2)e^{\nu _{4}}\int \frac{e^{-\nu _{4}}%
\widetilde{Z}_{(A)}^{(+)}}{(l-1)(l+2)r+6m}dr.
\]%
Following \ the relations (\ref{tilds}) we can compute the corresponding
''untileded'' values an put them in (\ref{auxc1}) in oder to find the
increments of fluctuations driven by the system of equations (\ref{peq1}).
For simplicity, we omit the rather compersome final expressions.

The formulas (\ref{peg3a}) together with a solution of the wave equation (%
\ref{peq3}) complete the procedure of definition of formal solutions for
polar perturbations. In Ref. \cite{chan} there are tabulated the data for
the potential (\ref{pot3}) for different values of $l$ and $\left(
l-1\right) (l+2)/2.$ In the anisotropic case the explicit form of solutions
is deformed by terms proportional to $\varepsilon n_{1}\sigma .$ The static
ellipsoidal like deformations can be modeled by the formulas obtained in the
limit $\sigma _{(\varepsilon )}\rightarrow 0.$

\section{The Stability of Black Ellipsoids}

The problem of stability of anholonomically deformed Schwarzschild metrics
to external perturbation is very important to be solved in order to
understand if such static black ellipsoid like objects may exist in general
relativity. In this context we address the question: Let be given any
initial values for a static locally anisotropic configuration confined to a
finite interval of $r_{\star },$ for axial perturbations, and $r_{\ast },$
for polar peturbations, will one remain bounded such peturbations at all
times of evolution?

We have proved that even for anisotropic configurations every type of
perturbations are governed by one dimensional wave equations of the form%
\begin{equation}
\frac{d^{2}Z}{d\rho }+\sigma ^{2}Z=VZ  \label{eq8}
\end{equation}%
where $\rho $ is a radial type coordinate, $Z$ is a corresponding $Z^{(\eta
)}$ or $Z_{(A)}^{(+)}$ with respective smooth real, independent of $\sigma
>0 $ potentials $\widetilde{V}^{(\eta )}$ or $V^{(-)}$ with bounded
integrals. For such equations a solution $Z(\rho ,\sigma ,\varphi _{0})$
satisfying the boundary conditions%
\begin{eqnarray*}
Z &\rightarrow &e^{i\sigma \rho }+R(\sigma )e^{-i\sigma \rho }\quad (\rho
\rightarrow +\infty ), \\
&\rightarrow &\qquad T(\sigma )e^{i\sigma \rho }\qquad (\rho \rightarrow
-\infty )
\end{eqnarray*}%
(the first expression corresponds to an incident wave of unit amplitude from
$+\infty $ giving rise to a reflected wave of amplitude $R(\sigma )$ at $%
+\infty $ and the second expression is for a transmitted wave of \ amplitude
$T(\sigma )$ at $-\infty ),$ provides a basic complete set of wave functions
which allows to obtain a stable evolution. For any initial perturbation that
is smooth and confined to finite interval of $\rho ,$ we can introduce the
integral%
\[
\psi (\rho ,0)=\left( 2\pi \right) ^{-1/2}\int_{-\infty }^{+\infty }\widehat{%
\psi }(\sigma ,0)Z(\rho ,\sigma )d\sigma
\]%
and define, at later times, the evoluiton of perturbations,%
\[
\psi (\rho ,t)=\left( 2\pi \right) ^{-1/2}\int_{-\infty }^{+\infty }\widehat{%
\psi }(\sigma ,0)e^{i\sigma t}Z(\rho ,\sigma )d\sigma .
\]%
The Schrodinger theory garantees the conditions%
\[
\int_{-\infty }^{+\infty }|\psi (\rho ,0)|^{2}d\rho =\int_{-\infty
}^{+\infty }|\widehat{\psi }(\sigma ,0)|^{2}d\sigma =\int_{-\infty
}^{+\infty }|\psi (\rho ,0)|^{2}d\rho ,
\]%
from which the boundedness of $\psi (\rho ,t)$ follows for all $t>0.$

In our consideration we have replaced the time partial derivative $\partial
/\partial t$ by $i\sigma ,$ which was represented by the approximation of
perturbations to be periodic like $e^{i\sigma t}.$ This is connected with a
time--depending variant of (\ref{eq8}), like%
\[
\frac{\partial ^{2}Z}{\partial t^{2}}=\frac{\partial ^{2}Z}{\partial \rho
^{2}}-VZ.
\]%
Multiplying this equation on $\partial \overline{Z}/\partial t,$ where $%
\overline{Z}$ denotes the complex conjugation, and integrating on parts, we
obtain
\[
\int_{-\infty }^{+\infty }\left( \frac{\partial \overline{Z}}{\partial t}%
\frac{\partial ^{2}Z}{\partial t^{2}}+\frac{\partial Z}{\partial \rho }\frac{%
\partial ^{2}\overline{Z}}{\partial t\partial \rho }+VZ\frac{\partial
\overline{Z}}{\partial t}\right) d\rho =0
\]%
providing the conditions of convergence of necessary integrals. This
equation added to its complex conjugate results in a constant energy
integral,
\[
\int_{-\infty }^{+\infty }\left( \left| \frac{\partial Z}{\partial t}\right|
^{2}+\left| \frac{\partial Z}{\partial \rho }\right| ^{2}+V\left| Z\right|
^{2}\right) d\rho =const,
\]%
which bounds the expression $|\partial Z/\partial t|^{2}$ and excludes an
exponential growth of any bounded solution of the equation (\ref{eq8}). We
note that this property holds for every type of ''ellipsoidal'' like
deformation of the potential, $V\rightarrow V+\varepsilon V^{(1)},$ with
possible dependences on polarization functions as we considered in (\ref%
{eq7a}) and/or (\ref{pot3}).

The general properties of the one--dimensional Schrodinger equations related
to perturbations of holonomic and anholonomic solutions of the vacuum
Einstein equations allow us to conclude that there are locally anisotropic
static configuratios which are stable under linear deformations.

In a similar manner we may analyze perturbations (axial or polar) governed
by a two--dimensional Schrodinger wayve equation like
\[
\frac{\partial ^{2}Z}{\partial t^{2}}=\frac{\partial ^{2}Z}{\partial \rho
^{2}}+A(\rho ,\varphi ,t)\frac{\partial ^{2}Z}{\partial \varphi ^{2}}-V(\rho
,\varphi ,t)Z
\]%
for some functions of necessary smooth class. The stability in this case is
proven if exists an (energy) integral
\[
\int_{0}^{\pi }\int_{-\infty }^{+\infty }\left( \left| \frac{\partial Z}{%
\partial t}\right| ^{2}+\left| \frac{\partial Z}{\partial \rho }\right|
^{2}+\left| A\frac{\partial Z}{\partial \rho }\right| ^{2}+V\left| Z\right|
^{2}\right) d\rho d\varphi =const
\]%
which bounds $|\partial Z/\partial t|^{2}$ $\ $for two--dimensional
perturbations. For simplicity, we omitted such calculus in this work.

Finally, we note that this way we can also prove the stability of
perturbations along ''anisotropic'' directions of arbitrary anholonomic
deformations of the Schwarzschild solution which have non--spherical
horizons and can be covered by a set of finite regions approximated as
small, ellipsoid like, deformations of some spherical hypersurfaces. We may
analyze the geodesic congruence on every deformed sub-region of necessary
smoothly class and proof the stability as we have done for the resolution
ellipsoid horizons. In general, we may consider horizons of with
non--trivial topology, like vacuum black tori, or higher genus anisotropic
configurations. This is not prohibited by the principles of topological
censorship \cite{ptc} if we are dealing with off--diagonal metrics and
associated anholonomic frames \cite{v}. The vacuum anholonomy in such cases
may be treated as an effective matter which change the conditions of
topological theorems.

\section{Outlook and Conclusions}

It is a remarkable fact that, in spite of appearance complexity, the
perturbations of static off--diagonal vacuum gravitational configurations
are governed by symilar types of equations as for diagonal holonomic
solutions. The origin of this mystery is located in the fact that by
anholnomic transforms we effectively diagonalized the off--diagonal metrics
by ''elongating'' some partial derivatives. This way the type of equations
governing the perturbations is preserved but, for small deformations, the
systems of linear equations for fluctuations became ''slightly'' nondiagonal
and with certain tetradic modifications of partial derivatives and
differentials. In details, the question of relating of particular integrals
of such systems associated with systems of linear differential equations is
investigated in Ref. \cite{chan}. In our case one holds the same relations
between the potentials $\widetilde{V}^{(\eta )}$ and $V^{(-)}$ and wave
functions $Z^{(\eta )}$ and $Z_{(A)}^{(+)}$ with that difference that the
physical values and formulas where polarized by some anisotropy functions $%
\eta _{3}(r,\theta ,\varphi ),\Omega (r,\varphi ),q(r),\eta (r,\varphi ),$ $%
w_{1}(r,\varphi )$ and $n_{1}(r,\varphi )$ and deformed on a small parameter
$\varepsilon .$

We also observe that the ''anisotropic'' potentials $\widetilde{V}^{(\eta )}$
and $V^{(-)}$ may be defined for such polarizations $\eta _{3}(r,\theta
,\varphi )$ as they would be smooth functions, integrable over the range of $%
r_{\ast },(-\infty ,+\infty )$ and positive everywhere. For real $\sigma ,$
the anisotropic solutions, represent ingoing and outgoing waves of type $%
e^{\pm i\sigma r_{\ast }}(r_{\ast }\rightarrow \pm \infty ).$ We conclude
that the underlying physical problem is one of reflexion and transmission of
incident waves (from $+$ or $-$ $\infty )$ by some anisotropically deformed
one--dimensional potential barriers $\widetilde{V}^{(\eta )}$ and/or $%
V^{(-)},$ for every fixed anisotropic angle $\varphi _{0}.$ Because at
distancies fare away from deformed horizons, the metrics of anisotropic
black holes transforms into the usual Schwarzschild solution we can prove
the equality of the reflexion and the transmission coefficients for the
axial and the polar perturgations, even the wave equations are
anholonomically deformed. Such formulas are tabulated in Ref. \cite{chan}
and may be used for anisotropic solutions but with some redefined
coefficients. The general properties of the one--dimensional
potential--scattering are preserved for small anholonomic deformations; they
are concerned with some solutions of one dimensional Schrodinger's wave
equations.

We emphasize that we proved the stability of black ellipsoids by fixing any
anisotro\-pic directions in the off--diagonal metrics, $\varphi =\varphi
_{0} $ i. e. under certain anholonomic constraints imposed on vacuum
gravitational configurations. A more general consideration with variable $%
\varphi $ relates the problem to the two dimensional Schrodinger's
potential--scattering problem, as well to an anholonomic Newman--Penrose
formalism which makes the solution of the problem of stability to be more
sophisticate. Nevertheless, we are having strong arguments that the
stability of anisotropic static solutions proved in the parametric
approximation $\varphi =\varphi _{0}$ will hold at least for such
polarizations which model two dimensional scattering effects containing some
particular one dimensional stable Schrodinger's potentials. The
one--dimensional perturbation analysis is the first, very important, step in
investigating the stability of any physical system; it should be included
into the higher dimensional and/or less constrained approaches.

We found that the origin of the results that the functions $Z^{(-)}$ and $%
Z^{(+)}$ and their anisotropic extensions $Z^{(\eta )}$ and $Z_{(A)}^{(+)}$
do in fact satisfy very similar one--dimensional wave equations steems from
the fact that some deep properties of static solutions of the Einstein
equations which hold both for diagonal and off--diagonal static--stationary
vacuum and/or electrovacuum metrics.

Finally, we conclude that there are static black ellipsoid vacuum
configurations which are stable with respect to one dimensional
perturbations, axial and/or polar ones, governed by solutions of the
corresponding one--dimensional Schrodinger equations. The problem of
stability of such objects with respect to two, or three, dimensional
perturbations, and the possibility of modeling such perturbations in the
framework of a two--, or three--, dimensional inverse scattering problem is
a topic of our further investigations.

\bigskip 

\subsection*{Acknowledgements}

~~ The work is supported by a NATO/Portugal fellowship at CENTRA, the
Instituto Superior Tecnico, Lisbon.

\section*{Appendix}

We compute the coefficients of the curvature tensor as%
\[
R_{\beta \gamma \tau }^{\alpha }=\delta _{\gamma }\Gamma _{\beta \tau
}^{\alpha }-\delta _{\tau }\Gamma _{\beta \gamma }^{\alpha }+\Gamma _{\sigma
\gamma }^{\alpha }\Gamma _{\beta \tau }^{\sigma }-\Gamma _{\sigma \tau
}^{\alpha }\Gamma _{\beta \gamma }^{\sigma },
\]%
of the Ricci tensor as
\[
R_{\beta \gamma \alpha }^{\alpha }=R_{\beta \gamma }
\]%
and of the Einstein tensor as%
\[
G_{\beta \gamma }=R_{\beta \gamma }-\frac{1}{2}g_{\beta \gamma }R
\]%
for $R=g^{\beta \gamma }R_{\beta \gamma }$ and $\delta _{\gamma }=\partial
_{\gamma }$ for $\gamma =2,3,4$ and $\delta _{1}=\partial _{1}-n_{1}\partial
/\partial t.$ Straightforward computations for the quadratic line element (%
\ref{metric2}) give%
\begin{eqnarray*}
R_{11} &=&-e^{-2\mu _{1}}[\delta _{11}^{2}(\mu _{3}+\mu _{4}+\mu
_{2})+\delta _{1}\mu _{3}\delta _{1}(\mu _{3}-\mu _{1})+\delta _{1}\mu
_{2}\delta _{1}(\mu _{2}-\mu _{1})+ \\
&&\delta _{1}\mu _{4}\delta _{1}(\mu _{4}-\mu _{1})]-e^{-2\mu _{2}}[\partial
_{22}^{2}\mu _{1}+\partial _{2}\mu _{1}\partial _{2}(\mu _{3}+\mu _{4}+\mu
_{1}-\mu _{2})]+ \\
&&e^{-2\mu _{4}}[\partial _{44}^{2}\mu _{1}+\partial _{4}\mu _{1}\partial
_{4}(\mu _{3}-\mu _{4}+\mu _{1}+\mu _{2})]-\frac{1}{2}e^{2(\mu _{3}-\mu
_{1})}[e^{-2\mu _{2}}Q_{12}^{2}+e^{-2\mu _{4}}Q_{14}^{2}], \\
R_{12} &=&-e^{-\mu _{1}-\mu _{2}}[\partial _{2}\delta _{1}(\mu _{3}+\mu
_{2})-\partial _{2}\mu _{1}\delta _{1}(\mu _{3}+\mu _{1})-\delta _{1}\mu
_{2}\partial _{4}(\mu _{3}+\mu _{1}) \\
&&+\delta _{1}\mu _{3}\partial _{2}\mu _{3}+\delta _{1}\mu _{4}\partial
_{2}\mu _{4}]+\frac{1}{2}e^{2\mu _{3}-2\mu _{4}-\mu _{1}-\mu
_{2}}Q_{14}Q_{24}, \\
R_{31} &=&-\frac{1}{2}e^{2\mu _{3}-\mu _{4}-\mu _{2}}[\partial _{2}(e^{3\mu
_{3}+\mu _{4}-\mu _{1}-\mu _{2}}Q_{21})+\partial _{4}(e^{3\mu _{3}-\mu
_{4}+\mu _{2}-\mu _{1}}Q_{41})], \\
R_{33} &=&-e^{-2\mu _{1}}[\delta _{11}^{2}\mu _{3}+\delta _{1}\mu _{3}\delta
_{1}(\mu _{3}+\mu _{4}+\mu _{2}-\mu _{1})]- \\
&&e^{-2\mu _{2}}[\partial _{22}^{2}\mu _{3}+\partial _{2}\mu _{3}\partial
_{2}(\mu _{3}+\mu _{4}-\mu _{2}+\mu _{1})]+\frac{1}{2}e^{2(\mu _{3}-\mu
_{1}-\mu _{2})}Q_{12}^{2}+ \\
&&e^{-2\mu _{4}}[\partial _{44}^{2}\mu _{3}+\partial _{4}\mu _{3}\partial
_{4}(\mu _{3}-\mu _{4}+\mu _{2}+\mu _{1})]-\frac{1}{2}e^{2(\mu _{3}-\mu
_{4})}[e^{-2\mu _{2}}Q_{24}^{2}+e^{-2\mu _{1}}Q_{14}^{2}], \\
R_{41} &=&-e^{-\mu _{1}-\mu _{4}}[\partial _{4}\delta _{1}(\mu _{3}+\mu
_{2})+\delta _{1}\mu _{3}\partial _{4}(\mu _{3}-\mu _{1})+\delta _{1}\mu
_{2}\partial _{4}(\mu _{2}-\mu _{1}) \\
&&-\delta _{1}\mu _{4}\partial _{4}(\mu _{3}+\mu _{2})]+\frac{1}{2}e^{2\mu
_{3}-\mu _{4}-\mu _{1}-2\mu _{2}}Q_{12}Q_{34}, \\
R_{43} &=&-\frac{1}{2}e^{2\mu _{3}-\mu _{1}-\mu _{2}}[\delta _{1}(e^{3\mu
_{3}-\mu _{4}-\mu _{1}+\mu _{2}}Q_{14})+\partial _{2}(e^{3\mu _{3}-\mu
_{4}+\mu _{1}-\mu _{2}}Q_{24})], \\
R_{44} &=&-e^{-2\mu _{4}}[\partial _{44}^{2}(\mu _{1}+\mu _{2}+\mu
_{3})+\partial _{4}\mu _{3}\partial _{4}(\mu _{3}-\mu _{4})+\partial _{4}\mu
_{1}\partial _{4}(\mu _{1}-\mu _{4})+ \\
&&\partial _{4}\mu _{2}\partial _{4}(\mu _{2}-\mu _{4})]+e^{-2\mu
_{1}}[\delta _{11}^{2}\mu _{4}+\delta _{1}\mu _{4}\delta _{1}(\mu _{3}+\mu
_{4}-\mu _{1}+\mu _{2})]+ \\
&&e^{-2\mu _{2}}[\partial _{22}^{2}\mu _{4}+\partial _{2}\mu _{4}\partial
_{2}(\mu _{3}+\mu _{4}-\mu _{1}+\mu _{2})]-\frac{1}{2}e^{2(\mu _{3}-\mu
_{4})}[e^{-2\mu _{1}}Q_{14}^{2}+e^{-2\mu _{2}}Q_{24}^{2}],
\end{eqnarray*}%
where the rest of coefficients are defined by similar formulas with a
corresponding changings of indices and partial derivative operators, $%
R_{22}, $ $R_{42}$ and $R_{32}$ is like $R_{11},R_{41}$ and $R_{31}$ with
with changing the index $1\rightarrow 2.$ The values $Q_{ij}$ and $Q_{i4}$
are defined respectively%
\[
Q_{ij}=\delta _{j}q_{i}-\delta _{i}q_{j}\mbox{ and }Q_{i4}=\partial
_{4}q_{i}-\delta _{i}\omega .
\]

The nontrivial coefficients of the Einstein tensor are
\begin{eqnarray*}
G_{11} &=&e^{-2\mu _{2}}[\partial _{22}^{2}(\mu _{3}+\mu _{4})+\partial
_{2}(\mu _{3}+\mu _{4})\partial _{2}(\mu _{4}-\mu _{2})+\partial _{2}\mu
_{3}\partial _{2}\mu _{3}]- \\
&&e^{-2\mu _{4}}[\partial _{44}^{2}(\mu _{3}+\mu _{2})+\partial _{4}(\mu
_{3}+\mu _{2})\partial _{4}(\mu _{2}-\mu _{4})+\partial _{4}\mu _{3}\partial
_{4}\mu _{3}]+ \\
&&e^{-2\mu _{1}}[\delta _{1}\mu _{4}+\delta _{1}(\mu _{3}+\mu _{2})+\delta
_{1}\mu _{3}\delta _{1}\mu _{2}]- \\
&&\frac{1}{4}e^{2\mu _{3}}[e^{-2(\mu _{1}+\mu _{2})}Q_{12}^{2}-e^{-2(\mu
_{1}+\mu _{4})}Q_{14}^{2}+e^{-2(\mu _{2}+\mu _{3})}Q_{24}^{2}], \\
G_{33} &=&e^{-2\mu _{1}}[\delta _{11}^{2}(\mu _{4}+\mu _{2})+\delta _{1}\mu
_{4}\delta _{1}(\mu _{4}-\mu _{1}+\mu _{2})+\delta _{1}\mu _{2}\delta
_{1}(\mu _{2}-\mu _{1})]+ \\
&&e^{-2\mu _{2}}[\partial _{22}^{2}(\mu _{4}+\mu _{1})+\partial _{2}(\mu
_{4}-\mu _{2}+\mu _{1})+\partial _{2}\mu _{1}\partial _{2}(\mu _{1}-\mu
_{2})]- \\
&&e^{-2\mu _{4}}[\partial _{44}^{2}(\mu _{1}+\mu _{2})+\partial _{4}\mu
_{1}\partial _{4}(\mu _{1}-\mu _{4})+\partial _{4}\mu _{2}\partial _{4}(\mu
_{2}-\mu _{4})+\partial _{4}\mu _{1}\partial _{4}\mu _{2}]+ \\
&&\frac{3}{4}e^{2\mu _{3}}[e^{-2(\mu _{1}+\mu _{2})}Q_{12}^{2}-e^{-2(\mu
_{1}+\mu _{4})}Q_{14}^{2}-e^{-2(\mu _{2}+\mu _{3})}Q_{24}^{2}], \\
G_{44} &=&e^{-2\mu _{1}}[\delta _{11}^{2}(\mu _{3}+\mu _{2})+\delta _{1}\mu
_{3}\delta _{1}(\mu _{3}-\mu _{1}+\mu _{2})+\delta _{1}\mu _{2}\delta
_{1}(\mu _{2}-\mu _{1})]- \\
&&e^{-2\mu _{2}}[\partial _{22}^{2}(\mu _{3}+\mu _{1})+\partial _{2}(\mu
_{3}-\mu _{2}+\mu _{1})+\partial _{2}\mu _{1}\partial _{2}(\mu _{1}-\mu
_{2})]-\frac{1}{4}e^{2(\mu _{3}-\mu _{1}-\mu _{2})}Q_{12}^{2} \\
&&+e^{-2\mu _{4}}[\partial _{4}\mu _{3}\partial _{4}(\mu _{1}+\mu
_{2})+\partial _{4}\mu _{1}\partial _{4}\mu _{2}]-\frac{1}{4}e^{2(\mu
_{3}-\mu _{4})}[e^{-2\mu _{1}}Q_{14}^{2}-e^{-2\mu _{2}}Q_{24}^{2}].
\end{eqnarray*}%
The component $G_{22}$ is to be found from $G_{11}$ by changing the index $%
1\rightarrow 2.$



\end{document}